# Mobile Application for GBAS Air Traffic Status Unit

*Hiba Zaidi*
*Amity School of Engineering and Technology, Noida, India*
hibawasi90@yahoo.in

**Abstract:** At present, the Air Traffic Status Unit (ATSU) is a windows PC based application, which receives the status of GBAS station over Ethernet and displays on the screen. The objective of this project is to convert the PC based Application into Mobile application using Android OS.

## 1. INTRODUCTION

The Air Traffic Status Unit (ATSU) is a read-only device functioning as a status monitor. It connects to the existing airport Local Area Network (LAN) via Ethernet and receives the Composite Status Message from the GBAS. The GBAS must be configured to send data to the ATSU [IP address]. The ATSU cannot command, configure, or request data from the GBAS.

Through the Android application the status can be seen on the android mobile rather than only on the PC. The Air Traffic Status can be accessed from anywhere in the world by multiple stakeholders The Mobile technology allows the staff handling the ATSU to stay alert without being tied to a single location. Whether the staff is travelling, out on calls, working from outside or from home anywhere on the globe, mobile devices can help them keep in touch, be attentive, and make use of the alerts and alarms. Thus, it leads to great flexibility and increases the efficiency. It also avoids the errors that occur due to the Ethernet cable connections. Because with this the ATSU data would be directly transferred as a mobile application. It also displays the service alerts and the alarms.

## 2. EQUIPMENT AND SOFTWARE

The ATSU is physically a laptop personal computer or any device capable of running in the Windows® operating environment and connecting to a LAN. The software required is the ATSU application itself. The Windows based application is now being converted into an Android operating system which would be a mobile application.

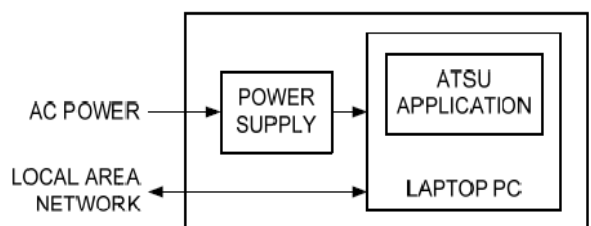
Figure 1 ATSU Subsystem Block Diagram [1]

### 2.1 Air Traffic Status Unit Operation
ATSU performs the following display functions:
- GBAS Mode: NORMAL, ALARM, TEST
- GLS Approach: AVAILABLE, PREDICTED OUTAGE, NOT AVAILABLE
- Count down timer to PREDICTED OUTAGE
- Current UTC date & time
- Connectivity to GBAS station

### 2.2 ATSU Display
The ATSU display when the GBAS Mode is Normal and the GLS Approach is Available is shown in Figure 2 ATSU Display, Normal & Available.

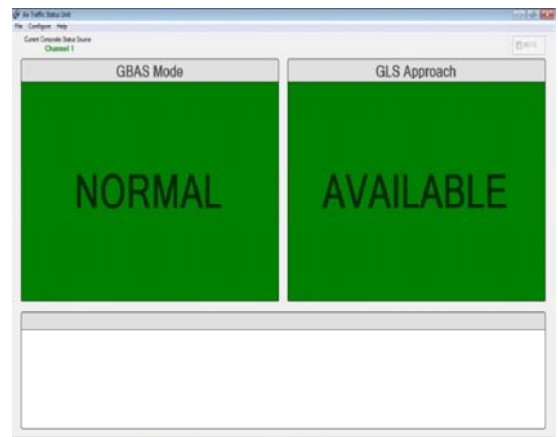
Figure 2 ATSU Display, NORMAL & AVAILABLE [1]

When the GBAS Mode is ALARM and the GLS Approach is NOT AVAILABLE the ATSU display appears as shown in Figure 3. The red color coding makes it clear that during this time, GBAS approach operations are not permitted.

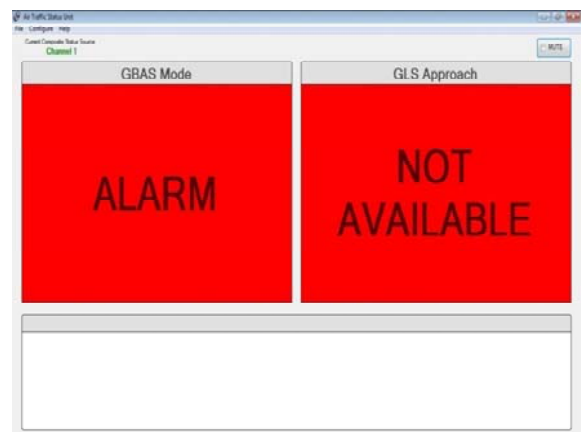
Figure 3 ATSU Display, ALARM & NOT AVAILABLE

GBAS TEST Mode and GLS Approach Not Available are depicted in Figure 4, again using red and also yellow to indicate that during this time, GBAS approach operations are not permitted of clicking these graphic figures.

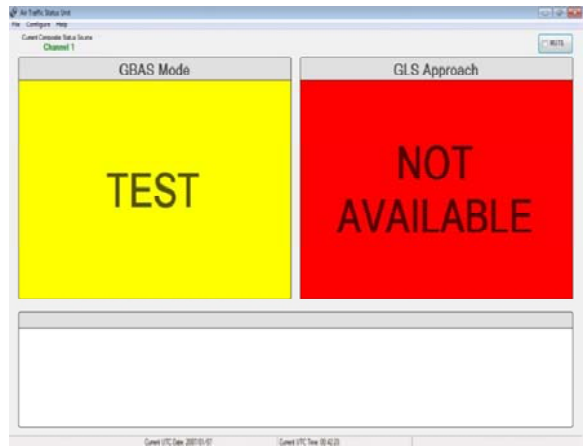

Figure 4 ATSU Display, TEST & NOT AVAILABLE

### 2.3 GBAS Operation in Predicted Outage

The GBAS can be operating normally and display Normal Mode when an outage has been predicted. The ATSU will display Predicted Outage in the yellow message block and the time to loss of service. Yellow means GLS approaches will soon be not available. While GLS Approach status remains Predicted Outage, aircraft may be cleared for landing as in normal operations. The display is organized to show the predicted start time of the outage and also countdown to the predicted start time.

### 2.4 GBAS Operation in Actual Constellation Alert

The GBAS can be operating normally and display Normal Mode during the actual outage. The ATSU will display *GLS Approach* as Not Available in the red message area block and the estimated return to service countdown timer. Red means GLS approaches are not available. Once the outage is over, the ATSU will display *GLS Approach* as Available. During the time of the outage, aircraft may not be cleared for approach. The display is organized to show the predicted end time of the outage and a countdown to the predicted end time.

### RESULTS

The mobile application for the Air Traffic Status Unit (ATSU) for GBAS was built similar to the existing PC application. The application displays the Status message, the Service Alerts and the Alarms in the Android device. The application in Android showing the splashscreen with the Honeywell logo (see Figure 5)

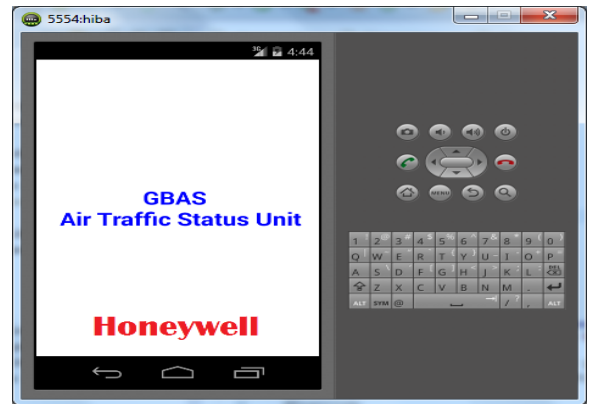

Figure 5 shows the splashscreen with Honeywell logo being created in Android

When no composite status message is available, the ATSU becomes grey (see Figure 6) to let the user know that there is no current data to be displayed.

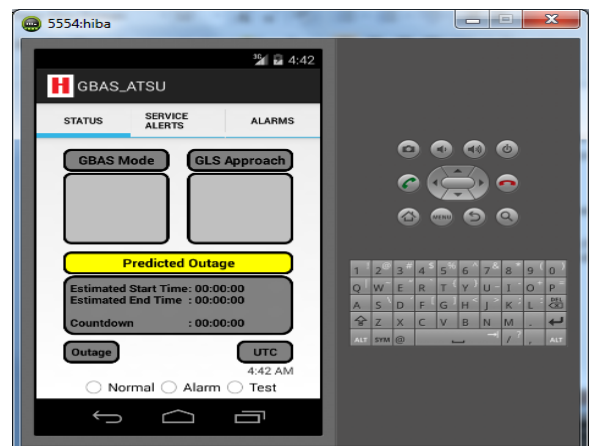

Figure 6 shows the image in Android with no status message

When the Normal radio button will be clicked the GABS Mode and GLS Approach becomes green in color. (see figure 7)

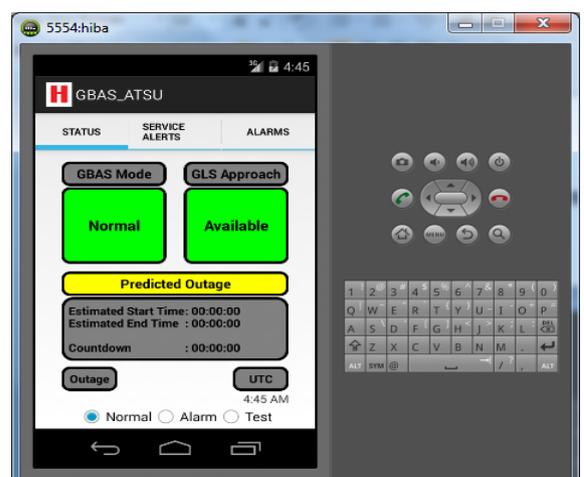

Figure 7 When the normal button is clicked

When the alarm button is clicked the GBAS Mode and the GLS Approach becomes Red. (see figure 8)

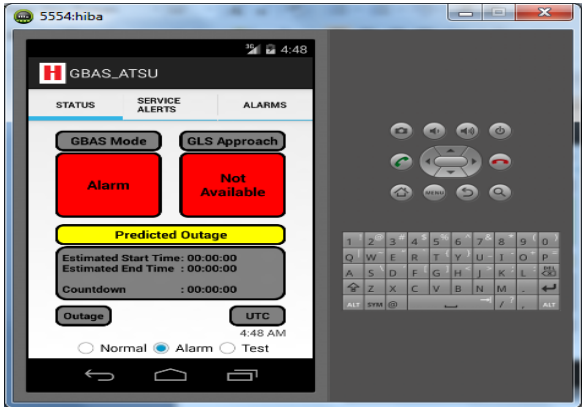
Figure 8 When the Alarm button is clicked

When the Test button is clicked the GBAS Mode becomes Red and the GLS Approach becomes Yellow in color (see figure 9)

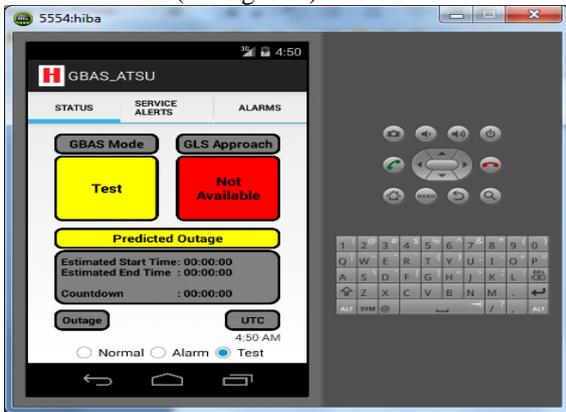
Figure 9 When the Test button is clicked

When we click on the outage button the Estimated Start Time that is 2 minutes more than the UTC Time comes up along with Estimated End Time. The countdown calculates the difference and it stops on getting 00:00:00.

Clicking on the other tab that is the Service Alerts will give all the 35 Alerts and their description. When we click on any alert and then press the display button the results comes up as a toast. (see figure 10)

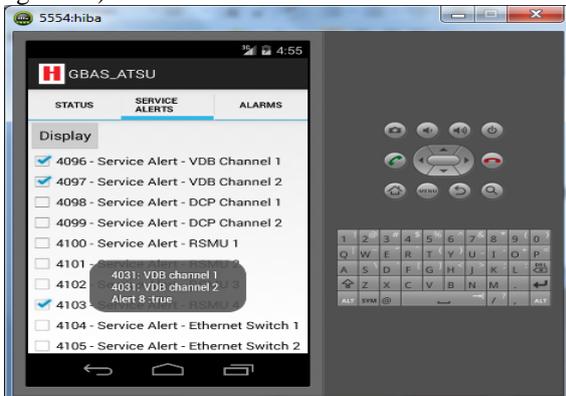
Figure 10 When the Service Alert tab is being clicked

Clicking on the other tab that is the Alarms will give all the 6 Alarms and their description. When we click on any alarm and then press the display button the results comes up as a toast. (see figure 11)

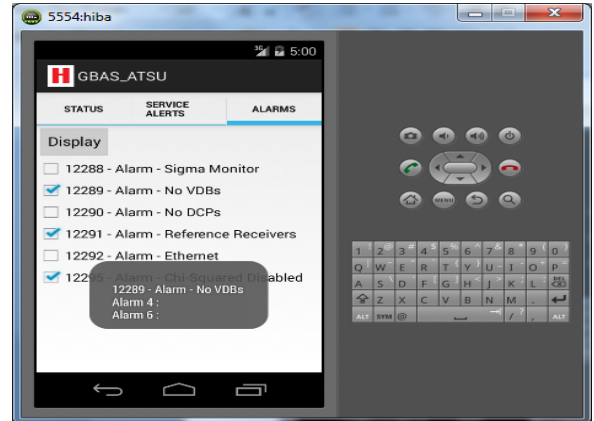
Figure 11 When the Alarm tab is clicked

## CONCLUSION

A mobile application for the Air Traffic Status Unit (ATSU) similar to the existing windows PC application was built using the Android operating system. The status message was displayed on the Android Application along with the Service Alerts and Alarms.

## REFERENCES

[1]"GBAS Commercial Information Book", Honeywell International INC. Aerospace – Minneapolis, MN USA

[3]Tim Murphy, and Thomas Imrich, "Implementation and Operational Use of Ground-Based Augmentation Systems (GBASs) VA Component of the Future Air Traffic Management System"Proceedings of the IEEE, Vol. 96, No. 12, December 2008.

[4]Irena Ambrožová1, and Stanislav Pleninger, "Implementation of GBAS System at the Václav Havel Airport", Number 2, Volume VIII, July 2013.

[5]"Performance-based NavigationHoneywell SmartPath$^{TM}$Ground Based Augmentation System (GBAS)",Pacific Aviation Directors WorkshopMarch 15, 2012

[6]Wei-Meng Lee, "Beginning Android 4 Application Development", John Wiley & Sons, Inc., Indianapolis, Indiana.